%% file: urls.tex
\documentclass[letterpaper,twocolumn,10pt]{article}
\usepackage{usenix,epsfig,endnotes}
\usepackage[square,sort,comma,numbers]{natbib}

\usepackage{graphicx,times}
\usepackage[hyphens]{url}
\usepackage{balance}
\usepackage{color}
\usepackage{multirow}
\usepackage[sort]{natbib} 
\usepackage{xspace}
\usepackage{listings}
\usepackage{float}
\restylefloat{table}
\usepackage[small,it]{caption}
\usepackage{tabularx}

\usepackage{dcolumn}
\makeatletter
\def\url@ulinkstyle{
  \@ifundefined{selectfont}{\def\UrlFont{\sf}}{\def\UrlFont{\small\ttfamily}}}
\makeatother
\usepackage[justification=centering]{caption}

\usepackage{pifont}% http://ctan.org/pkg/pifont
\newcommand{\cmark}{\ding{51}}%
\newcommand{\xmark}{\ding{55}}%

\usepackage{tikz}

\usepackage{multirow}

\usepackage[]{algorithm2e}

\newcommand{\paragraphbe}[1]{\vspace{0.75ex}\noindent{\bf \em #1} }

\begin{document}
\title{\Large \bf Gone in Six Characters: \\ 
Short URLs Considered Harmful for Cloud Services}

\author{
{\rm Martin Georgiev} \\
independent
\thanks{This research was done while the author was visiting Cornell Tech.}
\and
{\rm Vitaly Shmatikov}\\
Cornell Tech
}

\maketitle

\subsection*{Abstract}

Modern cloud services are designed to encourage and support collaboration.
To help users share links to online documents, maps, etc., several
services, including cloud storage providers such as Microsoft 
OneDrive\footnote{OneDrive was known as SkyDrive prior to January 27, 2014.} 
and mapping services such as Google Maps, directly integrate URL shorteners
that convert long, unwieldy URLs into short URLs, consisting
of a domain such as \url{1drv.ms} or \url{goo.gl} and a short token.

In this paper, we demonstrate that the space of 5- and 6-character
tokens included in short URLs is so small that it can be scanned using
brute-force search.  Therefore, all online resources that were intended
to be shared with a few trusted friends or collaborators are effectively
public and can be accessed by anyone.  This leads to serious security
and privacy vulnerabilities.

In the case of cloud storage, we focus on Microsoft OneDrive.  We show
how to use short-URL enumeration to discover and read shared content
stored in the OneDrive cloud, including even files for which the user
did not generate a short URL.  7\% of the OneDrive accounts exposed in
this fashion allow anyone to \emph{write} into them.  Since cloud-stored
files are automatically copied into users' personal computers and devices,
this is a vector for large-scale, automated malware injection.

% OneDrive APIs are designed to support easy programmatic access and allow
% traversal of users' accounts given a link to a single document.

In the case of online maps, we show how short-URL enumeration reveals the
directions that users shared with each other.  For many individual users,
this enables inference of their residential addresses, true identities,
and extremely sensitive locations they visited that, if publicly revealed,
would violate medical and financial privacy.

\section{Introduction}

Modern cloud services are designed to facilitate collaboration and
sharing of information.  To help users share links to online resources,
several popular services directly integrate URL shortening services that
convert long, unwieldy URLs into short URLs that are easy to send via
email, instant messages, etc.  For example, Microsoft OneDrive cloud
storage service uses the \url{1drv.ms} domain\footnote{When OneDrive
was SkyDrive, the domain for short URLs was \url{sdrv.ms}} for its short
URLs, Google Maps uses \url{goo.gl}, Bing Maps uses \url{binged.it}, etc.
In this paper, we investigate the security and privacy consequences of
this design decision.

First, we observe that the URLs created by many URL shortening services
are so short that the entire space of possible URLs can be scanned or
at least sampled on a large scale.  We then experimentally demonstrate
that such scanning is feasible.  Users who generate short URLs to their
online documents and maps may believe that this is safe because the
URLs are ``random-looking'' and not shared publicly.  Our analysis and
experiments show that these two conditions cannot prevent an adversary
from automatically discovering the true URLs of the cloud resources
shared by users.  Each resource shared via a short URL is thus effectively
\emph{public} and can be accessed by anyone anywhere in the world.

Second, we analyze the consequences of sharing for the users of cloud
storage services, using Microsoft OneDrive as our case study.  Like many
similar services, OneDrive (1) provides Web interfaces and APIs for easy
online access to cloud-stored files, and (2) automatically synchronizes
files between users' personal devices and cloud storage.  We demonstrate
that the discovery of a short URL for a single file in the user's OneDrive
account can expose \emph{all} other files and folders owned by the same
user and shared under the same capability key or without a capability
key\textemdash even files and folders that cannot be reached directly
through short URLs.  

% The ability to enumerate shared files and folders starting from a single
% URL would have been problematic even if the shared, short URLs had been
% drawn from a larger space, because a user who shares a single file with
% another user may not necessarily intend to reveal other shared files
% or folders in their account, let alone expose these files or folders to
% anyone on the Internet.

Because of ethical concerns, we did not download and analyze the
content of personal files exposed in this manner, but we argue
that OneDrive accounts are vulnerable to automated, large-scale
privacy breaches by less scrupulous adversaries who are not
constrained by ethics and law.  Recent compromises of Apple's cloud
services\footnote{\url{https://en.wikipedia.org/wiki/ICloud_leaks_of_celebrity_photos}}
demonstrated that users store very sensitive personal information in
their cloud storage accounts, sometimes intentionally and sometimes
accidentally due to automatic synchronization with their mobile phones.

% the combination of (a) URLs that are discoverable by brute-force
% enumeration and (b) programmatic account traversal from a single URL
% is deadly.

More than 7\% of OneDrive and Google Drive accounts we discovered by
scanning short URLs contain world-writable folders.  This means that an
adversary can automatically inject malicious content into these accounts.
Since the types of all shared files in an exposed folder are visible, the
malicious content can be format-specific, for example, macro viruses for
Word and Excel files, scripts for images, etc.  Furthermore, the adversary
can simply add executable files to these folders.  Because storage
accounts are automatically synchronized between the cloud and the user's
devices, this vulnerability becomes a vector for automated, large-scale
malware injection into the local systems of cloud-storage users.

Third, we analyze the consequences of public sharing for the users
of online mapping services such as Google Maps, MapQuest, Bing Maps,
and Yahoo!\ Maps.  Short-URL enumeration reveals not only the locations
that users shared with each other, but also \emph{directions} between
locations.  In many cases, these directions start from or terminate
at single-family residential addresses and allow inference of users'
identities via cross-correlation with public directories such as White
Pages.  In addition, residential-to-residential directions could reveal
the existence of personal relationships, including those intended to
remain discreet.  Even worse, many of the destinations mapped by users
are highly sensitive, including hospitals, clinics, and physicians
associated with specific diseases (e.g., mental illnesses and cancer)
or procedures (e.g., abortion); correctional and juvenile detention
facilities; places of worship; pawnbrokers, payday and car-title loan
stores, etc.  Analytics APIs can also be invoked on individual maps to
reveal the exact time when the directions were obtained and how often
the map was referred to, thus providing further context.

In summary, our analysis shows that automatically generated short URLs
are a terrible idea for cloud services.  When a service generates a URL
based on a 5- or 6-character token for an online resource that one user
wants to share with another, this resource effectively becomes public
and universally accessible.  Combined with other design decisions,
such as Web APIs for accessing cloud-stored files and retrieving user-
or resource-specific metadata, as well as automatic synchronization of
files and folders between personal devices and cloud storage, universal
public access to online resources leads to significant security and
privacy vulnerabilities.

\section{Background}

\subsection{URL Shorteners}

Uniform Resource Locators (URLs) are the standard method for addressing
Web content.  URLs often encode session management and/or document
structure information and can grow to hundreds of characters in length.
The HTTP standard~\cite{rfc2616} does not specify an a priori limit on
the length of a URL, but implementations impose various restrictions,
limiting URLs to 2048 characters in practice~\cite{URIlength}.

Long URLs are difficult to distribute and remember.  When printed on
paper media, they are difficult to read and type into the browser.
Even when shared via electronic means such as email and blog posts,
long URLs are not elegant because they are often broken into multiple
lines.  The problem is exacerbated when the URL contains (URL-encoded)
special characters, which may be accidentally modified or filtered out
by sanitization code aiming to block cross-site scripting and injection
attacks.  Another motivation for URL shortening comes from services
like Twitter that impose a 140-character limit on the messages users
post online and from mobile SMS that are limited to 160 characters,
making it impossible to share long URLs.

URL shortening services (URL shorteners) map long URLs to short ones.
The first URL shorteners were patented in 2000~\cite{patent}.  Hundreds
of URL shorteners have emerged on the market since then~\cite{list}.
Many services offer additional features such as page-view counting,
analytics for tracking page visitors' OS, browser, location, and referrer
page, URL-to-QR encoding, etc.

% Users have gotten comfortable using URL shorteners, even though many
% do not know how URL shortening works, nor understand the security and
% privacy implications of short URLs.  As a result, users rely on short
% URLs even when sharing access to private Web content such as personal
% pictures and documents with friends, family, and colleagues.

\begin{figure}[!htp]
\vspace{2ex}
\centering
\includegraphics[scale=0.3]{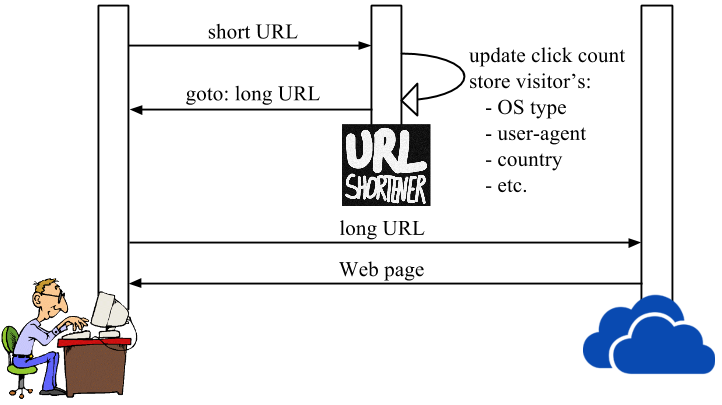}
\captionsetup{justification=raggedright}
\caption{Resolving short URLs.}
\label{fig:url-resolution}
\end{figure}

A URL shortener accepts a URL as input and generates a short URL.
The service maintains an internal database mapping each short URL to
its corresponding original URL so that any online access using a short
URL can be resolved appropriately (see Figure~\ref{fig:url-resolution}).

To generate short URLs, URL shorteners first define the alphabet (most
commonly, [a-z,A-Z,0-9]) and the length of the output \emph{token}.
The token, sometimes referred to as the \emph{key}, is the last part
of the short URL, differentiating individual links in the shortener's
internal database.  For example, if the alphabet is [a-z,A-Z,0-9] and
the token is 6 characters long, the shortener can generate $62^6 \approx
5.7 \cdot 10^{10}$ possible short URLs.

Short URLs can be generated sequentially, randomly, using a combination
of the two (as in the case of \url{bit.ly}~\cite{secpriv}), or by hashing
the original URL.  Sequential generation reveals the service's usage
patterns and introduces concurrency issues.

% \begin{algorithm}
%  \KwData{long URL, alphabet, token length}
%  \KwResult{short URL}
%    \If{long URL is in database}{
%    \Return short URL $\leftarrow$ database[long URL]
%    }
%    number = -1\\
%    MAX\_NUMBER = alphabet size$^{token length}$ - 1\\
%    \While{number is in database}{
%      number $\leftarrow$ Random(MAX\_NUMBER)\\  
%    }
%    short URL $\leftarrow$ URL shortener's URL prefix + GetToken(number)\\
%    database[number] $\leftarrow$ long URL to short URL mapping\\
%  \Return short URL\\
%  \caption{Short URL generation}
% \end{algorithm}

\url{bit.ly} is a popular URL shortener.  According to the counter on the
front page of \url{bitly.com}, the company claims to have shortened over
26 billion URLs at the time of this writing.  The tokens in \url{bit.ly}
URLs are between 4 and 7 characters long, but currently the first
character in 7-character tokens is almost always 1, thus the effective
space of 7-character \url{bit.ly} URLs is $62^6$ as described above.
Therefore, the overall space of \url{bit.ly} URLs is $62^{4} + 62^{5}
+ 2 \cdot 62^{6} \approx 1.2 \cdot 10^{11}$.

Some cloud services integrate URL shortening into their products to help
users share links.  For example, Microsoft OneDrive uses \url{1drv.ms}
for this purpose.  Reverse DNS lookup shows that \url{1drv.ms} is a
branded short domain~\cite{branded} operated by \url{bit.ly}.  Therefore,
OneDrive short URLs are in effect \url{bit.ly} short URLs.  This fact
has two implications: (1) \url{bit.ly} and \url{1drv.ms} share the same
token space; (2) \url{1drv.ms} URLs can be resolved by the \url{bit.ly}
resolver.  Note that \url{bit.ly} URLs cannot be resolved using the
\url{1drv.ms} resolver unless they point to OneDrive documents.

Other branded domains operated by \url{bit.ly} include \url{binged.it} for
Bing Maps, \url{yhoo.it} for Yahoo!\ Maps, and \url{mapq.st} for MapQuest.
All of them currently use 7-character tokens with the first character
set to 1.

Google Maps uses the \url{goo.gl/maps} domain and, prior to the changes
made in response to this paper (see Section~\ref{disclosure}), 5-character
tokens.  Thus, the entire token space of \url{goo.gl/maps} was $62^{5}
\approx 9.2 \cdot 10^{8}$.

\subsection{Cloud Storage Services}

Cloud storage services are gaining popularity because they enable users
to access their files from anywhere and automatically synchronize files
and folders between the user's devices and his or her cloud storage.

\subsubsection{OneDrive}

OneDrive is an online cloud storage service operated by Microsoft.
The first 5 GB of storage are free; larger quotas are available for a
small monthly fee.

%an additional 100 GB can be purchased
%for \$1.99 or 200 GB for \$3.99 per month.  Users who purchase access
%to Microsoft's Office 365 can receive 1 TB of storage for \$9.99 a month.

OneDrive currently allows Word, Excel, PowerPoint, PDF, OneNote,
and plain-text files to be viewed and edited through the service's Web
interface.  OneDrive also supports online viewing of many image and video
file formats, such as JPEG, PNG, MPEG etc.  Users may share OneDrive
files and folders with view-only, edit, and public-access capabilities.

OneDrive provides client applications for Mac, PC, Android, iOS, Windows
Phone, and Xbox to facilitate automatic file and folder synchronization
between user's devices and his cloud storage account.

To facilitate application development and programmatic access to OneDrive
accounts, Microsoft distributes two different, independent SDKs: Live
SDK~\cite{api} and OneDrive pickers and savers SDK~\cite{pickers}.
Live SDK is built using open standards like OAuth 2.0, REST, and JSON.
It supports full-fledged access to files, folders, albums, photos, videos,
audio files, tags, and comments.  The lightweight OneDrive pickers and
savers SDK supports limited functionality such as opening and storing
OneDrive files and creating links to shared files.

\subsubsection{Google Drive}

Google Drive is Google's cloud storage product.  New users
get 15GB of storage for free; larger quotas, similar to OneDrive,  
are available for a small fee.

% with the option to buy 100GB for \$1.99,
%or 1TB for \$9.99.  Higher storage amounts are also available: 10TB
%(\$99.99), 20TB (\$199.99), or 30TB (\$299.99).

Google Drive has built-in support for Docs, Sheets, Slides, Forms,
Drawings, and Maps.  Users can thus view and edit popular file types like
DOC, DOCX, PPT, PPTX, XLS, XLSX, etc.  Users can also install applications
from Google's Web Store that extend Google Drive's functionality to
specialized file formats such as PhotoShop's PSD and AutoCAD's DWG.

Google Drive provides client applications for Mac, PC, Android, and iOS
which automatically synchronize files and folders between the user's
devices and his or her cloud storage account.

To facilitate programmatic access to files and folders stored on
Google Drive, Google provides Google Drive SDKs~\cite{gdrivesdk} for
Android, iOS, and the Web.   Additionally, Google Drive API v.2 is
available~\cite{gdriveapi}.

\subsection{Online Mapping Services}

Online maps are among the most popular and essential cloud-based services.
MapQuest offered Web-based maps in 1996, followed by Yahoo!\ Maps in 2002,
Google Maps in 2005, and Bing Maps in 2010.  In addition to driving
directions, modern online maps provide traffic details, road conditions,
satellite, bird's-eye, and street views, 3D imagery of notable locations,
etc.

All online mapping services let users share locations, as well as driving
directions between two or more locations.  The corresponding URLs are
very long, thus mapping services directly integrate URL shorteners
into their user interfaces, helping users share maps via text messages,
social media, and email.

Mapping services provide APIs and SDKs to application developers.  Google
Maps~\cite{googlemapsapi} distributes SDKs for Android, iOS, and the Web.
Bing Maps provides an SDK for Windows Store apps~\cite{bingmapssdk}
and AJAX and REST APIs for Web and mobile~\cite{bingmapsapi}.
There is also an unofficial, community-supported Bing Maps Android
SDK~\cite{bingmapsandroidsdk}.  MapQuest supports Web Services,
JavaScript, and Flash APIs~\cite{mapquestapi}.  Yahoo!\ discontinued their
Yahoo!\ Maps Web Services in 2011, but previously they had provided Flash,
AJAX, and Map Image APIs~\cite{yahoomapsapi}.

\section{Scanning Short URLs}

\paragraphbe{Scanning rates.}
\url{bit.ly} provides an API~\cite{bitly-api} for querying its database.
Access to this API is currently rate-limited to five concurrent
connections from a single client, with additional ``per-month,
per-hour, per-minute, per-user, and per-IP rate limits for each API
method''~\cite{ratelimiting}.  The limits are not publicly disclosed.
When a limit is reached, the API method stops processing further
requests from the client and replies with HTTP status code 403.  In our
experiments, a simple, unoptimized client can query the \url{bit.ly}
database at a sustained rate of 2.6 queries/second over long periods of
time.  Further optimizations may push the effective query rate closer to
the stated 5 queries/second rate limit and sustain it over a long time.
We also observed that much higher rates, up to 227 queries/second, are
possible for brief periods before the client's IP address is temporarily
blocked by \url{bit.ly}.

\url{goo.gl/maps} also provides an API~\cite{googleapi} for
querying its database.  The free usage quota is 1,000,000 queries per
day~\cite{googleapiquota}.  At the time of our experiments, there was
also an option to request a higher quota.

\paragraphbe{Sampling.}
To generate random tokens for the 6-character and 7-character token space
of \url{bit.ly} and the 5-character token space of \url{goo.gl/maps},
we first defined the alphabet: [a-z,A-Z,0-9].  We then calculated the
maximum number that a token can represent when interpreted as a Java
BigInteger~\cite{bigint}
%\begin{center}
%\emph{BigInteger maxNum = new BigInteger(alphabetSize);}
%\end{center}
%We obtained a new random number generator:
%\begin{center}
%\emph{Random prng = new Random();}
%\end{center}
and generated a random number within this space, interpreting it as a token.
%\begin{center}
%\emph{BigInteger randNum = new BigInteger(64, prng);}\\
%\emph{randNum = randNum.mod(maxNum);}
%\end{center}
Random tokens in our samples were generated without replacement.
The process of token generation ran until the desired number of unique
random tokens was obtained for each target service (e.g., \url{bit.ly})
and target token space (e.g., 6-character token space.)

To sample the space of \url{bit.ly} URLs, we generated 100,000,000 random
6-character tokens and queried \url{bit.ly} from 189 machines.  Our sample
constitutes 0.176\% of the 6-character token space.  We found 42,229,055
URL mappings.  Since the query tokens were chosen randomly, this implies
that the space of 6-character \url{bit.ly} URLs has approximately 42\%
density.  Because not all characters in \url{bit.ly} URLs appear to be
random~\cite{secpriv}, there exist areas of higher density that would
yield valid URLs at an even higher rate.

We also randomly sampled the 7-character token space on \url{bit.ly}.
At the time of our experiments, \url{bit.ly} set the first
character in all\footnote{With a few hard linked exceptions like
\url{http://bit.ly/BUBVDAY}} 7-character tokens to 1.  Thus, in
practice, the search space of 7-character tokens has the same size as
the space of 6-character tokens.  Similarly to the 6-character scan,
we generated 100,000,000 random tokens by setting the first character to
1 and appending a randomly generated 6-character token.  The resulting
sample constituted 0.176\% of the 7-character token space and produced
29,331,099 URL mappings.  Thus, the space of 7-character \url{bit.ly}
URLs has approximately 29\% density.

A careful reader will notice that if our density estimates are correct,
\url{bit.ly} must have shortened more than 0.42 $\cdot$ 62$^{6}$ + 0.29
$\cdot$ 62$^{6}$ $\approx$ 40 billion URLs.  Yet, the counter on the
front page of \url{bitly.com} says that they shortened 26 billion URLs.
We conjecture that this discrepancy is due to some URLs (e.g., those
under branded domains) not being counted towards the reported total.

\url{goo.gl/maps} has a much smaller token space: $9.2 \cdot 10^{8}$
vs.\ $1.2 \cdot 10^{11}$.  Prior to changes made by Google in response to
our report (see Section~\ref{disclosure}), we scanned 63,970,000 tokens
$\approx$ 7\% of the entire token space.  Our scan produced 23,965,718
URL mappings, implying that the density on \url{goo.gl/maps} is 37.5\%.

\paragraphbe{Exhaustive enumeration.}
At the current effective rate of querying \url{bit.ly}, enumerating
the entire \url{bit.ly} database would take approximately 12.2 million
compute hours, roughly equivalent to 510,000 client-days.  Amazon EC2
Spot Instances~\cite{spotinstances} may be a cost-effective resource
for automated URL scanning.  Spot Instances allow bidding on spare
Amazon EC2 instances, but without guaranteed timeslots.  The lack of
reserved timeslots matters little for scanning tasks.  At the time we
were conducting our scanning experiments, Amazon EC2 Spot Instances cost
\$0.003 per hour~\cite{spotprices}, thus scanning the entire \url{bit.ly}
URL space would have cost approximately \$36,700.  This price will drop
in the future as computing resources are constantly becoming cheaper.
Moreover, Amazon AWS offers a free tier~\cite{awsfree} service to new
users with 750 free micro-instance hours of Linux plus 750 micro-instance
hours of Windows per month for 12 months.  Therefore, a stealthy attacker
who is able to register hundreds of new AWS accounts can enumerate the
entire \url{bit.ly} database for free.

Prior to changes described in Section~\ref{disclosure}, enumerating the
entire \url{goo.gl/maps} database would have required 916 client-days.
Google Cloud Platform offers a \$300 credit~\cite{googlefree} to be used
over 60 days.  Therefore, a stealthy attacker capable of registering a few
hundred Google accounts could have enumerated the entire \url{goo.gl/maps}
database for free in a matter of hours.

\section{Short URLs in Cloud Storage Services}

Cloud storage services create a unique URL for each file and folder
stored in the user's account.  These URLs allow users to view and edit
individual files via the Web interface, change the metadata associated
with files and folders, and share files and folders with other users.

Sharing actual URLs is often inconvenient: email agents may wrap long
URLs, rendering them unclickable, text messages and Twitter have a limit
on message size, etc.  URL shortening helps users share URLs over email,
text or instant messages, and social media.

\subsection{Microsoft OneDrive}

The experiments in this section used short-URL scanning to discover
publicly accessible OneDrive files and folders.  Our scanner accessed
only public URLs and did not circumvent any access-control protections.
Information was collected solely for measurement purposes.

Our scanner considered only the \emph{metadata}, such as files and
directory names.  We did not analyze the contents of OneDrive files
found by scanning because they may contain sensitive personal data.
Note that these contents remain exposed through public URLs and are thus
vulnerable to a less scrupulous adversary.

%\mg{The account owner's email address is not available from the metadata,
%only their first + last name.}

\begin{table*}[!htb]
\scalebox{0.94}{
  \centering
  \begin{tabular}{| l | c | l | c | c | c | c | c | c | c | }
    \hline
    \textbf{File type}  & \textbf{prefix}                                                               & \textbf{path}         &\textbf{cid}   & \textbf{id}   & \textbf{resid}        & \textbf{app}  & \textbf{v} & \textbf{ithint} & \textbf{authkey}       \\ \hline
    Word                        & \multirow{9}{*}{\url{https://onedrive.live.com}}              & /view$^{*}$           & \cmark                & \xmark                & \cmark                & \cmark                & \xmark        & \xmark                & \multirow{9}{*}{optional}\\ \cline{1-1}\cline{3-9}
    Excel                       &                                                                               & /view or /edit                & \cmark                & \xmark                & \cmark                & \cmark                & \xmark   & \xmark             & \\ \cline{1-1}\cline{3-9}
    PowerPoint          &                                                                               & /view$^{*}$           & \cmark                & \xmark                & \cmark                & \cmark                & \xmark        & \xmark                & \\ \cline{1-1}\cline{3-9}
    OneNote             &                                                                               & /view or /edit                & \cmark                & \xmark                & \cmark                & \cmark                & \xmark        & \xmark                & \\ \cline{1-1}\cline{3-9}
    PDF                 &                                                                               & /view                 & \cmark                & \xmark                & \cmark                & \cmark                & \xmark        & \xmark                & \\ \cline{1-1}\cline{3-9}
    Surveys             &                                                                               & /survey                       & \xmark                & \xmark                & \cmark                & \xmark                & \xmark        & \xmark                & \\ \cline{1-1}\cline{3-9}
    Media files         &                                                                               & /                             & \cmark        & \cmark                & \xmark                & \xmark                & \cmark        & \xmark                & \\ \cline{1-1}\cline{3-9}
    Downloads           &                                                                               & /download.aspx        & \cmark                & \xmark                & \cmark                & \xmark                & \xmark        & \xmark                & \\ \cline{1-1}\cline{3-9}
    Folders                     &                                                                               & / $^{+}$                      & \cmark                & \cmark                & \xmark                & \xmark                & \xmark        & \cmark                & \\ \hline
  \end{tabular}
}
\captionsetup{justification=raggedright}
\caption{OneDrive URL formats.\\$^{*}$ Word and PowerPoint files shared with ``edit'' capability can be edited online, despite the absence of ``/edit'' path.\\$^{+}$ Folders shared with ``edit'' allow anyone to write into them. }
  \label{t:urlformats}
\end{table*}

\subsubsection{Discovering OneDrive Accounts}

Of the 42,229,055 URLs we discovered from the 6-character token space of
\url{bit.ly}, 3,003 URLs (0.003\% of the sample space) reference files
or folders under the \url{onedrive.live.com} domain.  Additionally,
16,521 URLs (0.016\% of the sample space) reference files or folders
under the \url{skydrive.live.com} domain.  If this density holds over the
entire space, the full scan would produce $62^{6} \cdot 0.003\% \approx
1,700,000$ (respectively, $62^{6} \cdot 0.016\% \approx 9,000,000$) URLs
pointing to OneDrive (respectively, SkyDrive) documents.  In our sample
scan, each client found, on average, 43 OneDrive/SkyDrive URLs per day.
At this rate, it would take approximately 245,000 client-days to enumerate
all OneDrive/SkyDrive URLs mapped to 6-character tokens.  A botnet can
easily achieve this goal in a single day or even much faster if the
operator is willing to have bots' IP addresses blocked by \url{bit.ly}.

Of the 29,331,099 URLs we discovered from the 7-character token space of
\url{bit.ly}, 25,594 (0.025\% of the sample space) point to
OneDrive files or folders, and 21,487 (0.021\% of the sample
space) point to SkyDrive files or folders.  Thus, the projected URL counts
of OneDrive/SkyDrive links in the 7-character token space of \url{bit.ly} 
are $62^{6} \cdot 0.025\% \approx 14,200,000$, and 
$62^{6} \cdot 0.021\% \approx 11,900,000$, respectively.

For each OneDrive/SkyDrive URL found by our sample scan, the scanner
issued a GET request.  If the landing page did not redirect to a page
outside the user's account, we considered the link ``live.'' The number
of live links is generally greater than the number of OneDrive/SkyDrive
accounts because different links may lead to different files in the
same account.

Of the 3,003 OneDrive URLs (respectively, 16,521 SkyDrive URLs) sampled
from the 6-character token space, 2,130 (respectively, 9,694) were live.
Of the 25,594 OneDrive URLs (respectively, 21,487 SkyDrive URLs) sampled
from the 7-character token space, 22,069 (respectively, 13,472) were live.

All URLs in our sample lead to distinct OneDrive accounts.  Due to the
small sample size, we cannot draw any conclusions about the total number
of OneDrive accounts that would be discovered by a full scan.

\subsubsection{Traversing OneDrive Accounts}
\label{traverse}

OneDrive supports all URL formats shown in Table~\ref{t:urlformats}.
Each account is uniquely identified by the value of the \emph{cid}
parameter.  The \emph{id} and \emph{resid} parameters have the
``\emph{cid!sequence\_number}'' format.  Thus, given \emph{id} or
\emph{resid}, it is trivial to recover \emph{cid}, but given \emph{cid},
there is no easy way to construct a valid \emph{id} or \emph{resid}.
However, these sequence numbers can be brute-forced.  Possible values for
the \emph{app} parameter are \emph{Word}, \emph{Excel}, \emph{PowerPoint},
\emph{OneNote}, and \emph{WordPdf}.  We observed only the value of
\emph{3} for the \emph{v} parameter.  The \emph{ithint} parameter denotes
a folder and encodes the type of content therein, such as JPEG PNG,
or PDF.  The \emph{authkey} parameter is a capability key that grants
access rights (view-only, edit, etc.)

It is not necessary to guess URL parameter values to gain access to
OneDrive files.  Having obtained the URL of a single document, one
can exploit the predictable structure of OneDrive URLs to traverse the
account's directory tree and enumerate other shared files and folders.
The account traversal methodology described in the rest of this section
worked reliably between October 2014 and February 2016.  As of March
2016, direct access to the account's root URL (see below) no longer
reveals the URLs of files and folders shared under the same capability
in that account.

Suppose a scan found a short URL such as
\url{http://1drv.ms/1xNOWV7} which resolves to
\url{https://onedrive.live.com/?cid=485bef1a805} \url{39148&id=485BEF1A80539148!115&}
\url{ithint=folder,xlsx&authkey=!AOOp2TqTTSMT5} \url{q4}.  Parse this URL
and extract the \emph{cid} and \emph{authkey} parameters, then construct
the root URL for the account by replacing \emph{XXX} and \emph{YYY} in
\url{https://onedrive.live.com/?cid=XXX&authkey=YYY} with the \emph{cid}
and \emph{authkey} values.

\begin{table*}[!htp]
\scalebox{1.14}{
  \centering
  \begin{tabular}{| l | r | r |}
    \hline
    \textbf{File type}  & \textbf{\# of files found in 6-char sample space} & \textbf{\# of files found in 7-char sample space}  \\ \hline
    Word                        & 2,116   & 21,077      \\ \hline
    Excel                       & 921     & 6,050       \\ \hline
    PowerPoint                  & 688     & 5,068       \\ \hline
    OneNote                     & 51      & 6           \\ \hline
    PDF                         & 10,080  & 41,465      \\ \hline
    Surveys                     & 22      & 226         \\ \hline
    Media files                 & 204,735 & 862,641     \\ \hline
    Downloads$^{*}$             & 8,663   & 168,613     \\ \hline
  \end{tabular}
}
\captionsetup{justification=raggedright}
  \caption{Publicly accessible files on OneDrive discovered by sampling from the 6- and 7-character token space of bit.ly.
  \\$^{*}$ ``Downloads'' refers to file types not natively supported for viewing or editing via the OneDrive Web interface.}
  \label{t:filetypes}
\end{table*}

Prior to March 2016, access to the root URL made it easy to automatically
discover URLs of shared files and folders in the account.  For example,
to find URLs of individual files, parse the HTML code of the page and look
for ``\emph{a}'' elements with ``\emph{href}'' attributes containing
``\emph{\&app=}'', ``\emph{\&v=}'', ``\emph{/download.aspx?}'',
or ``\emph{/survey?}''. Such links point to individual documents.
Links that start with \url{https://onedrive.live.com/} and contain the
account's \emph{cid} may lead to other folders.

Starting from each of the 2,130 OneDrive URLs discovered by our sample
scan of the 6-character token space of \url{bit.ly} and navigating
through the directory trees of the corresponding OneDrive accounts,
we found a total of 227,276 publicly accessible files.  Similarly,
navigating from each of the 22,069 OneDrive URLs discovered in the
7-character token space yielded a total of 1,105,146 publicly accessible
files (see Table~\ref{t:filetypes}).

\begin{figure}[!Ht]
\centering
\includegraphics[scale=0.35]{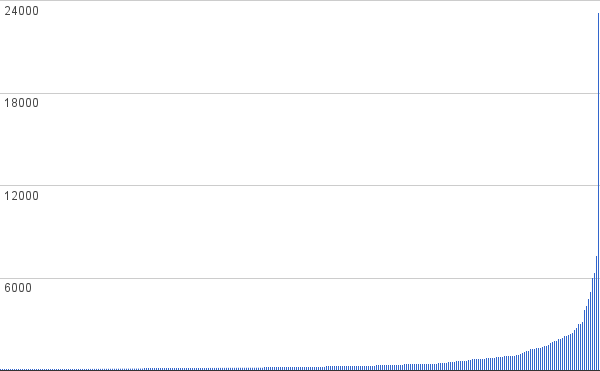}
\captionsetup{justification=raggedright}
\caption{Distribution of files per OneDrive account discovered by scanning
the 6-character token space of bit.ly}
%\vspace{0.5em}
\label{fig:filedist6}
\vspace{1em}
\end{figure}

\begin{figure}[!Ht]
\centering
\includegraphics[scale=0.35]{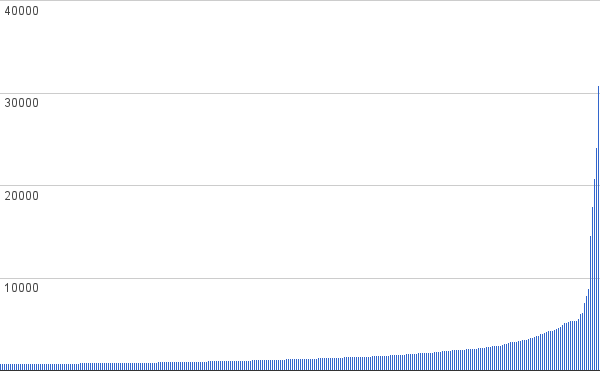}
\captionsetup{justification=raggedright}
\caption{Distribution of files per OneDrive account discovered by scanning
the 7-character token space of bit.ly}
%\vspace{0.5em}
\label{fig:filedist7}
\vspace{1em}
\end{figure}

Figure~\ref{fig:filedist6} shows the distribution of files per account in
our sample of the 6-character token space.  The average number of files
per account is 106, the maximum is 23,240, the minimum is 0 (i.e., an
empty folder).  The distribution of files per account in the 7-character
token space is shown in Figure~\ref{fig:filedist7}.  The average is 50,
the maximum is 30,779, the minimum is 0.

\subsubsection{Exploiting Unlocked OneDrive Folders}

Among the 2,130 live OneDrive accounts discovered in our sample
of the 6-character token space of \url{bit.ly}, 150 have at least
one folder shared with edit functionality.  To find such accounts,
our scanner searched the HTML code of the page for a ``\emph{span}''
element with ``\emph{class}'' attribute equal to ``\emph{navLinkText}''
and text attribute equal to ``\emph{Upload}''.  Of the 22,069 OneDrive
accounts found in our sample of the 7-character token space, 1,561
have at least one folder shared with edit functionality.  We estimate
that approximately \textbf{7\%} of discoverable OneDrive accounts have
world-writable folders.

%\begin{figure}[!Ht]
%\centering
%\includegraphics[scale=0.35]{imgs/DistributionOfFilesPerAccount.png}
%\captionsetup{justification=raggedright}
%\caption{Distribution of files per OneDrive account discovered by scanning
%the 6-character token space of bit.ly}
%\label{fig:filedist6}
%\vspace{1em}
%\end{figure}

%\begin{figure}[!Ht]
%\centering
%\includegraphics[scale=0.35]{imgs/DistributionOfFilesPerAccount7.png}
%\captionsetup{justification=raggedright}
%\caption{Distribution of files per OneDrive account discovered by scanning
%the 7-character token space of bit.ly}
%\label{fig:filedist7}
%\vspace{1em}
%\end{figure}

We call these folders ``unlocked'' because \emph{anyone} who
knows their URL\textemdash which, as we demonstrated, is easily
discoverable\textemdash can use the edit feature to overwrite existing
files and/or add new files, potentially planting malware into users'
OneDrive accounts.  Microsoft appears to perform some rudimentary
anti-virus scanning on OneDrive accounts, but it is trivial to evade.
For example, this scanning fails to discover even the test EICAR
virus\footnote{The EICAR Standard Anti-Virus Test file is a special
'dummy' file used to check and confirm the correct operation of security
products.} compressed in the \texttt{.xz} format.

% Microsoft previously claimed not to scan files uploaded into OneDrive for
% viruses.\footnote{\url{http://answers.microsoft.com/en-us/onedrive/forum/sdfiles-sdother/are-the-files-on-onedrive-being-virus-scanned/09ceeb53-1dc2-4173-864f-1cef250fd413}}

Automatic synchronization between the OneDrive cloud and users' personal
machines and devices, which is normally a very convenient feature,
turns this vulnerability into a major security hole.  For example, if
the attacker infects a user's existing file (e.g., inserts a macro virus
into a Word or Excel file), all of the victim's devices linked to his
or her OneDrive account will automatically download the infected file.
When the victim opens the file, the malware will execute with the victim's
privileges on the machine where the file was opened.

The attacker can also add new files to unlocked folders, for example,
executable malware files with names designed to trick the user into
clicking on them when these files automatically appear on the user's
personal computer or device.  This attack vector can also be leveraged
to exploit any number of known bugs in parsers and renderers of common
file formats such as JPEG~\cite{jpegbugs}, PDF~\cite{pdfbugs}, and
DOCX~\cite{docxbugs}.

\subsection{Google Drive}

Unlike OneDrive, Google Drive does not directly integrate a URL shortener,
thus users need to manually invoke a shortener if they want to generate
a short URL.

Our sample scan of 6-character \url{bit.ly} tokens yielded 44 links
to Google Drive folders: 30 are view-only, 3 are writable, 7 have
already been taken down, and 4 are permission-protected.  Our sample
of 7-character tokens yielded 414 links to Google Drive folders: 277
are view-only, 40 are writable, 49 have been taken down, and 48 are
permission-protected.  As with OneDrive, anyone who discovers the URL
of a writable Google Drive folder can upload arbitrary content into it,
which will be automatically synced with the user's devices.

Unlike OneDrive, Google Drive allows access to documents that were
removed from the folders but not permanently deleted from the trash.

\section{Short URLs in Mapping Services}

In this section, we first show that by scanning short URLs, one can
discover driving directions shared by users of online mapping services.
We then explain how these directions compromise users' privacy by
revealing sensitive locations they visited, their social ties, etc.

Our analysis focuses on Google Maps because the token space of
\url{goo.gl/maps} URLs was so small prior to the changes described
in Section~\ref{disclosure}.  We believe that similar results can
be obtained for any mapping service that integrates URL shorteners,
including MapQuest, Bing Maps, and Yahoo!\ Maps.

\paragraphbe{Google Maps.}
Of the 23,965,718 URLs in our Google Maps sample, 2,357,844 or about 9.8\%
are for directions; the rest are for individual locations.

\begin{figure*}[!htp]
\minipage{0.5\textwidth}
\centering
\includegraphics[scale=0.2]{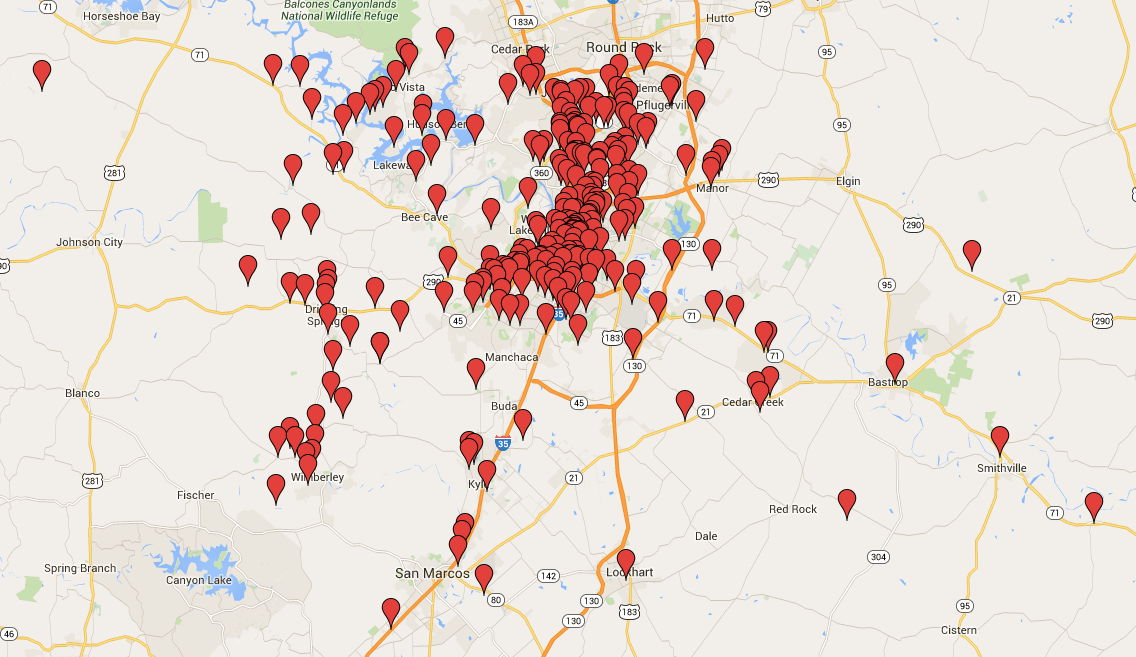}
\captionsetup{justification=raggedright}
\caption{Locations associated with a single user in Austin, TX.}
\label{fig:geocache}
\endminipage\hfill
\minipage{0.5\textwidth}
\centering
\includegraphics[scale=0.2]{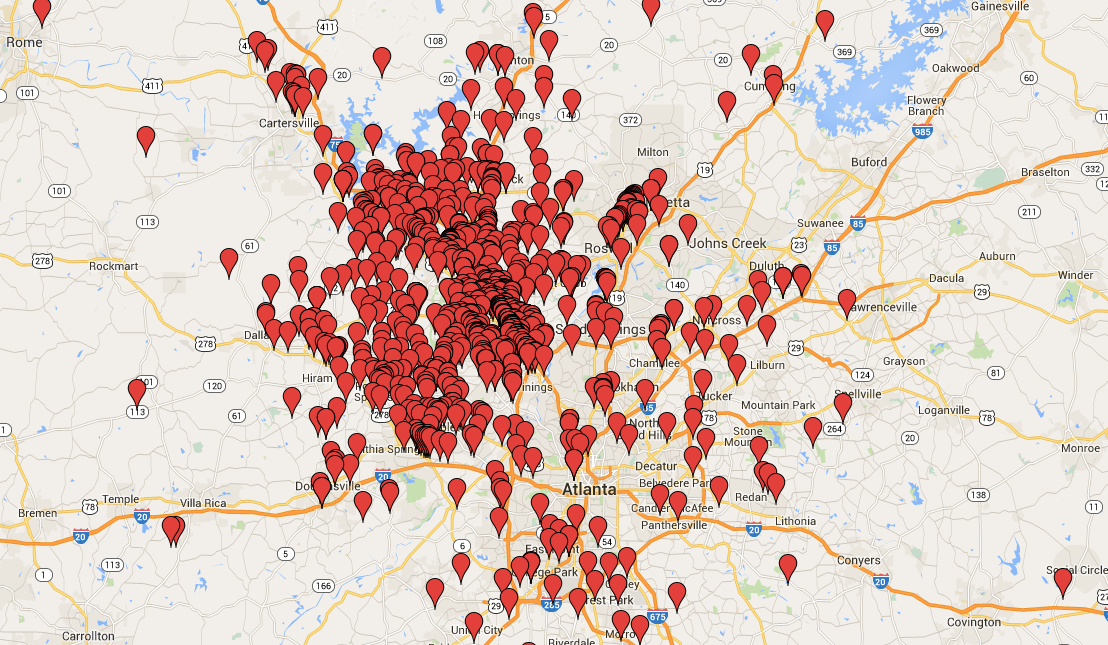}
\captionsetup{justification=raggedright}
\caption{Locations associated with D \& D Autows Inc.}
\label{fig:towingcompanymap}
\endminipage
\end{figure*}

\begin{figure}[!h]
\centering
\includegraphics[scale=0.44]{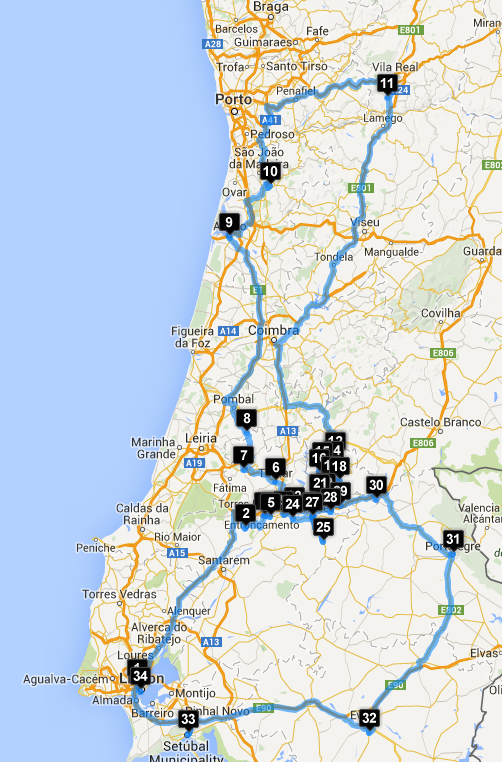}
\captionsetup{justification=raggedright}
\caption{Residential addresses associated with Hospital Doutor Manoel Const{\^{a}}ncio.}
\label{fig:map-graph}
\end{figure}

Our sample includes directions to and from many sensitive locations:
clinics for specific diseases (including cancer and mental illnesses),
addiction treatment centers, abortion providers, correctional and
juvenile-detention facilities, payday and car-title lenders, gentlemen's
clubs, etc.  In particular, the sample contains 3,913 map directions
that start at a hospital and end at a residential address, and 12,668
directions that start at a residential address and end at a hospital.
Figure~\ref{fig:map-graph} shows an example.  We could have constructed
similar maps for any sensitive location.  More importantly, \emph{anyone}
could have constructed them simply by scanning all \url{goo.gl/maps} URLs.

The endpoints of driving directions shared via short URLs often contain
enough information to uniquely identify the individuals who requested
the directions.  For instance, when analyzing one such endpoint, we
uncovered the address, full name, and age of a young woman who shared
directions to a planned parenthood facility.

Conversely, by starting from a residential address and mapping all
addresses appearing as the endpoints of the directions to and from the
initial address, one can create a map of who visited whom.

Fine-grained data associated with individual residential addresses
can be used to infer interesting information about the residents.
For instance, we conjecture that one of the most frequently occurring
residential addresses in our sample (see Figure~\ref{fig:geocache}) is
the residence of a geocaching enthusiast.  He or she shared directions
to hundreds of locations around Austin, TX, many of them specified as
GPS coordinates.  We have been able to find some of these coordinates
in a geocaching database~\cite{geocaching}.

Similarly, we can take a business (e.g., D \& D Autows Inc.) and
extract all map directions created to or from its location\textemdash
see Figure~\ref{fig:towingcompanymap}.  If a person had their vehicle
towed to or from their house, then their identity can be easily inferred
by cross-correlating the addresses with public directories such as
White Pages.

Additionally, Google API for short URLs reveals the exact time when
the URL was created, as well as the approximate time of recent URL
visits~\cite{urlanalytics}.  This information can be used to create
fine-grained activity profiles for users who share Google Maps directions.

\paragraphbe{MapQuest.}
MapQuest uses the branded \url{mapq.st} domain operated by \url{bit.ly}.
Our 6-character sample contains 151,334 short MapQuest URLs: 6,737 for
directions and 144,597 for point locations.  Our 7-character sample
contains 141,721 URLs: 66,929 for directions and 74,792 for point
locations.  All new short MapQuest URLs use 7-character \url{bit.ly}
tokens.

\paragraphbe{Bing Maps.}
Bing Maps uses the branded \url{binged.it} domain operated by
\url{bit.ly}.  Our 6-character sample contains 28,271 Bing Maps URLs:
10,020 for directions and 18,251 for point locations.  Our 7-character
sample contains 34,363 URLs: 13,116 for directions and 21,247 for point
locations.  All new short Bing Maps URLs use 7-character \url{bit.ly}
tokens.

\paragraphbe{Yahoo!\ Maps.}
Yahoo!\ Maps uses the branded \url{yhoo.it} domain operated by
\url{bit.ly}.  Our 6-character sample contains 1,331 Yahoo!\ Maps URLs:
771 for directions and 560 for point locations. Our 7-character sample
contains 8,400 URLs: 6,207 for directions and 2,193 for point locations.
All new short Yahoo!\ Maps URLs use 7-character \url{bit.ly} tokens.

\input{mitigation}

\section{Related Work}
Antoniades et al.~\cite{we-b} explored the popularity, temporal activity,
and performance impact of short URLs.

Neumann et al.~\cite{secpriv} studied the security and privacy
implications of URL shorteners on Twitter, focusing on the use of short
URLs for spam and user tracking, as well as leakage of private information
via URL-encoded parameters and HTTP referer headers.  As part of their
analysis, they discovered and manually examined 71 documents hosted on
Google Docs.

% (it is not clear whether they received the permission from the affected
% users to examine the private Google Docs they discovered).

In contrast to Neumann et al., we demonstrate that it is practically
feasible to automatically discover a large number of cloud-stored files by
randomly scanning short URLs.  We show how automated traversal of OneDrive
accounts reveals even files and folders that do not have short URLs
associated with them.  We also identify large-scale malware injection as a
serious security risk for cloud storage accounts discovered via short-URL
scanning.  Unlike Neumann et al., we did not analyze the content of
private documents we found due to ethical and legal considerations and the
logistical difficulties of requesting permission from the affected users.

Klien and Strohmaier~\cite{spam} investigated the use of short URLs for
fraud, deceit, and spam.  Based on the logs of the URL shortener their
group operated over several months, they concluded that about 80\% of
the URLs they shortened are spam-related.  This analysis does not apply
to short URLs integrated into cloud services.

Maggi et al.~\cite{threats} built a service for crowdsourced users to
preview the landing pages of short URLs resolved by 622 URL shorteners
and found that short URLs are often used to hide the true URLs of drive-by
download and phishing pages.  They also explored the countermeasures that
can be deployed by URL shorteners.  They did not discover the problem
of cloud storage accounts or mapping directions exposed by short URLs.

There is a rich literature on inferring information about individuals
from location data.  Becker et al.~\cite{ugroups} used anonymized
call detail records from a large US communications service provider
to identify large groups of people who collectively share the same
usage patterns.  Crandall et al.~\cite{socialties} inferred social ties
between people based on their co-occurrence in a geographic location.
Isaacman et al.~\cite{places} inferred important places in people's
lives from location traces.  Montjoye et al.~\cite{ucrowd} observed
that 95\% of individuals can be uniquely identified given only 4 points
in a high-resolution dataset such as a cell phone carrier's service
records.  Golle and Partridge~\cite{locpairs} showed the feasibility
of re-identifying anonymized location traces; futility of anonymizing
location traces was also demonstrated in~\cite{locanon,mtraces}.

Between 2013 and 2015, information about many Uber rides, including
customers' exact addresses, was accidentally made public after Google
indexed ``share your ETA'' links posted by Uber's customers~\cite{uber}.
Uber fixed the problem by expiring the links after 48 hours.

To the best of our knowledge, this paper is the first to observe that the
sharing of maps between users can lead to significant privacy violations
because short URLs integrated into popular mapping services effectively
make all shared locations and directions public.

\section{Conclusions}

URL shortening, which looks like a relatively minor feature, turns out
to have serious consequences for the security and privacy of modern
cloud services.  In this paper, we demonstrate that the token space of
short URLs is so small as to be efficiently enumerable and scannable.
Therefore, any short link to an online document or map shared by a user
of a cloud service is effectively public.

In the case of cloud-storage services such as Microsoft OneDrive, this
not only leads to leakage of sensitive documents, but also enables anyone
to inject arbitrary malicious content into unlocked accounts, which
is then automatically copied into all of the account owner's devices.

In the case of mapping services, short URLs reveal addresses
and\textemdash via easy cross-correlation with public
directories\textemdash  identities of users who shared directions
to medical facilities (including abortion, mental-health, and
addiction-treatment clinics), prisons and juvenile detention centers,
places of worship, and other sensitive locations; enable inference of
social ties between people; and leak other sensitive private information.

Solving the problem identified in this paper will not be easy since
short URLs are an integral part of many cloud services and previously
shared information remains publicly accessible (unless URL shorteners
take the drastic step of revoking all previously issued short URLs).
We present several recommendations which could mitigate the damage caused
by short URLs.

\section{Disclosure}
\label{disclosure}

\paragraphbe{Microsoft.}
We notified Microsoft about the security and privacy risks of short
OneDrive URLs on May 28, 2015.  In particular, any user who shares a
short OneDrive URL with a collaborator may unintentionally expose the
shared files and folders to \emph{everyone}.  Furthermore, if the shared
documents and folders allow writing, anyone can inject malicious content
into them that will be automatically downloaded to the user's computers
and devices.

After an email exchange involving several messages, ``Brian'' from
Microsoft's Security Response Center (MSRC) informed us on August 1,
2015, that the ability to share documents via short URLs ``appears by
design,'' and thus ``does not currently warrant an MSRC case.''

In March of 2016, Microsoft removed the
``shorten link'' option from OneDrive, causing a number of user
complaints.\footnote{http://answers.microsoft.com/en-us/onedrive/forum/odwork-odshare/shorten-link-option-no-longer-available-on/bc3dc4eb-cb54-43e0-bcff-a072e8dba3ad?auth=1}
We asked MSRC whether this change was made in response to our previous
report.  MSRC informed us that our analysis played no role in their
decision to remove this option and reiterated that they do not consider
our report a security vulnerability.  At approximately the same time,
Microsoft changed the API so that the account traversal methodology
described in Section~\ref{traverse} no longer appears to work.

As of this writing, all previously generated short OneDrive URLs remain
vulnerable to scanning and malware injection.

\paragraphbe{Google.} 
We notified Google about the privacy risks of short Google Maps URLs
on September 15, 2015.  Google promptly responded to our report.  As of
September 21, 2015, newly created short URLs to Google Maps have 11 or
12-character tokens and are thus not vulnerable to brute-force scanning.

\newpage

\balance

\begin{flushleft}
{\footnotesize 
    \bibliographystyle{acm}
    \bibliography{urls}
}
\end{flushleft}

\end{document}

%% file: mitigation.tex
\section{Mitigation}
\label{mitigation}

% In this section we overview potential mitigation strategies for services
% that provide and/or are paired with URL shortening.  One proposal,
% of course, would be to eliminate the use of short URLs entirely.
% While abandoning them could help security, it would also mean loosing
% the convenience and benefits discussed earlier in the paper.

We suggest five approaches to mitigate the vulnerabilities identified in
this paper: (1) make short URLs longer, (2) inform users about the risks
of URL shorteners, (3) do not rely on universal URL shorteners, (4) employ
CAPTCHAs or other methods to separate human users from automated scanners,
and (5) design better APIs for the cloud services that use short URLs.

First, URL shorteners should use longer URLs.  There is an obvious
tension between maintaining the benefits of short URLs and preventing
scanning attacks.  It might be instructive to compare the size of token
space to the ``bits of security'' metric sometimes used in cryptography.
Most token alphabets consist of $62$ characters, which is close to $2^6$.
Each character can thus be thought of as providing roughly $6$ bits of
search space.  We estimate that tokens of 10 characters or more would
make it difficult to scan the entire token space.  From a usability
perspective, 10-character tokens will slightly increase the difficulty
of hand-copying short URLs and also make them less ``friendly-looking.''
From an attack mitigation perspective, longer tokens would be highly
effective.  For example, at the current density of 42\% in the 6-character
token space of \url{bit.ly}, the attacker needs about 2 queries to obtain
a single valid URL.  Should the token size be increased to 10 characters,
the attacker would have to send 35 million queries to obtain a single
valid URL.  Future decisions on the size of token space should involve
careful analysis of attackers' capabilities.

Second, URL shorteners should warn users that creating a short URL may
expose the content behind the URL to third parties.  For integrated
applications, the warnings can be more specific and tailored to the
application (e.g., maps).  This approach has limitations.  A typical user
may not be able to properly assess whether using a shortener is dangerous.
Furthermore, the person who is asking for a URL to be shortened could be
different from the person who is impacted by its disclosure.  For example,
a towing company may not care about disclosing the residences it serviced
even if the individuals being serviced do.

Third, cloud services should consider using internal, company-owned URL
shorteners, as opposed to universal shorteners such as \url{bit.ly}.
This change would enable companies to (1) significantly decrease the
density of the token space, (2) closely monitor automated scans of
the short-URL space, and (3) take appropriate actions as soon as a scan is
detected.  Furthermore, it will increase the burden on the attackers since
they will need to scan different token spaces for different services.

Fourth, URL shorteners must take a more aggressive approach against
scanning.  Instead of fixed monthly/daily/hourly limits, they
should identify large-scale scanners and block their IP addresses.
Alternatively, they could ask users to solve CAPTCHAs every few hundred
requests to verify that the requester is human.

Fifth, cloud services that use URL shorteners need better API design.
Lengthening short URLs does not prevent an attacker who discovers a
short URL to a single file from enumerating all files and folders shared
under the same capability key.  In particular, Microsoft OneDrive should
change the format and structure of long URLs so that, given the URL of one
document, it is no longer possible to discover the URLs of other documents
in the same account (Microsoft appears to have made this change some
time between February and March of 2016).  A similar approach is already
taken by Google Drive when individual files are shared.